\documentclass[11pt]{article}
\usepackage[text={6.5in,8.75in},centering]{geometry}                
\usepackage{graphicx}
\usepackage{amssymb}
\usepackage{epstopdf}
\usepackage{amsmath}
\usepackage{amssymb}
\usepackage{latexsym}
\usepackage{float}
\usepackage{fancyheadings}

\newcommand{\ket}[1]{\left|#1\right\rangle}

\newcommand{\doeparticle}{\it This work is supported by the U.S. Department of Energy under cooperative research agreement Contract Number DE-FG02-05ER41360.}
\title{\uppercase{Fractional and Majorana Fermions}: \\ 
The Physics of Zero Energy Modes\footnote{Nobel Symposium, Stockholm, Sweden, May 2010} \ 
                             \footnote{Semat Lecture, CCNY, New York, NY, April 2011}}
\author{R. Jackiw\\
\it\small Center for Theoretical Physics\\
\it\small Massachusetts Institute of Technology\\
\it\small Cambridge, MA 02139}
\date{}
\begin{document}
\maketitle
\thispagestyle{fancy}
\begin{abstract}
We describe the occurrence and physical role of zero-energy modes in the Dirac equation with a topologically non-trivial background.
\end{abstract}
\newpage
\section*{Introduction}
 My goal is to describe new states of matter, which were initially encountered in purely theoretical investigations within the mathematical formalism used in physics, and subsequently were looked for and sometimes discovered in actual experiments. But 
first I want to remind you of an earlier instance of such a development advancing physics: 
Dirac's discovery of anti-matter. 

Dirac was seeking a relativistic and quantum mechanical equation to describe spin 1/2 electrons-free electrons to begin with. He proposed a matrix equation, first order in derivatives. Here it is.  
 \begin{figure}[h]
\includegraphics[scale=.25]{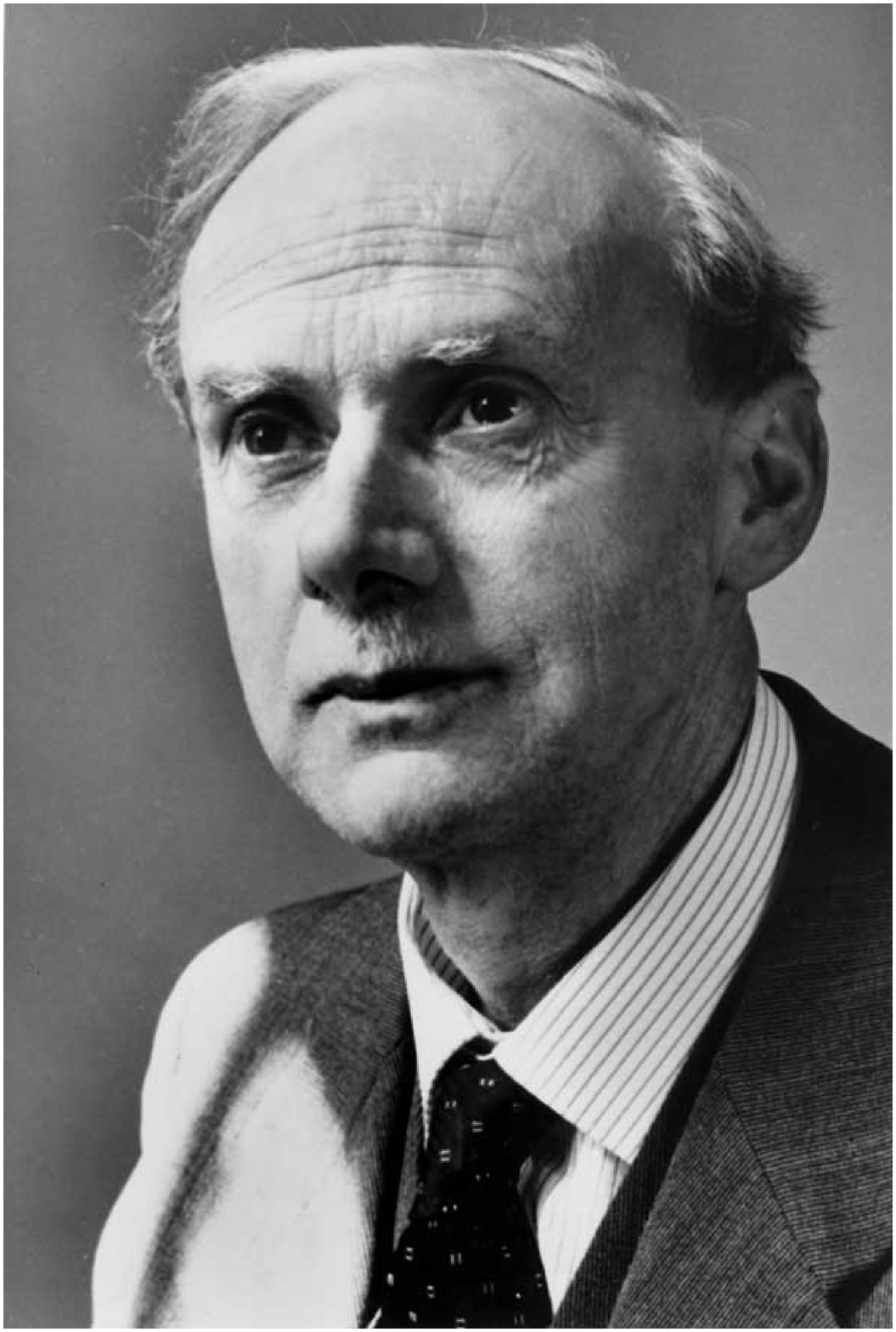} \quad
\begin{minipage}[b][3.25in][t]{4in}
{\underline{\bf Dirac Matrix Equation}}\\[2ex]
$({\boldsymbol \alpha} \cdot {\bf p} + \beta m)\, {\mathrm{\Psi}} = i\ \frac{\partial}{\partial t} \ \Psi$\\[1ex]
$\Psi$ complex (charged excitations)\\[1ex]
${\boldsymbol \alpha}, \beta \Rightarrow$ matrices, ${\bf p} \Rightarrow  \frac{1}{i}\ \nabla, m \Rightarrow$mass\\[1ex]
energy eigenvalues:\\
$({\boldsymbol \alpha} \cdot {\bf p} + \beta m)\ \Psi_E = E \Psi_E$\\
$E > 0 \quad$ empty in vacuum\\[1ex]
$E< 0 \quad$ filled in vacuum \\[1ex]
 $m$ produces gap\\[1ex]
\mbox{\hspace{.4in}vacuum charge vanishes}
\end{minipage}
\end{figure}

The Dirac equation is a beautiful equation of mathematical physics. It is also a beautiful equation in mathematics because, as Dirac himself stated, it provides a square root for the Laplacian of the Schr\"{o}dinger equation, and the Laplacian is a much studied mathematical operator.
 
But square roots come with both signs, and the energy eigenvalue $E$ of the Dirac operator comes out positive, which is good, but also negative, which requires reinterpretation, because physical electrons carry only positive energy.

With characteristic faith in the mathematical formalism, Dirac did not reject his equation with its negative energy modes. Rather he interpreted them as describing oppositely charged anti-electrons. This speculation was soon confirmed by the experimental discovery of positrons.
 Evidently keeping faith with a beautiful equation, more specifically, taking seriously the eigenvalue spectrum, no matter what its peculiarities may be, leads to physical discovery!

There is no more news about Dirac's original equation. Current research proceeds because his equation has been extended and deformed in various ways. With these extensions we have encountered further spectral peculiarities, which are mathematically fascinating and whose physical interpretation exposes novel excitations in physical systems.

Deformations of Dirac's equation arise from two motivations:  mathematical and physical. Mathematicians facing a beautiful equation instinctively look for the more general case, for an extended structure. On the other hand, physicists have established that Dirac-like equations describe low-energy excitations in actual condensed matter systems.  Such descriptions are of course approximate, but this does not detract from their usefulness in a restricted context. After all, even the original Dirac equation provides only an approximate treatment of relativistic electrons. 

Electron motion within materials extends over the volume  occupied by the sample. However in certain materials electron motion is effectively confined to a line or to a plane.    Correspondingly our Dirac equation needs to be reformulated on a one-dimensional line or on a two-dimensional plane, in addition  to the  three-dimensional volume.

Even before delving into details, we can appreciate how aptly the Dirac equation confronts condensed matter phenomena: its positive and negative energy eigenvalues model free conduction electrons and bound valence electrons respectively. In the ground state the positive energy modes are empty because there are no conduction electrons, while the negative energy modes are filled with valence electrons. The mass term in the equation produces by its absolute value a gap between the conduction and valence bands, as in an insulator; a vanishing mass term leads to a gapless spectrum, as in a conductor. 

 A profound deformation of the Dirac equation is achieved when the constant in space mass term is replaced by an inhomogenous term that varies with position. Only profiles that are significantly different from the homogenous case are interesting because small variations give rise to small, insignificant changes. To guide us in selecting interesting profiles, we classify them into distinct classes.  Such classification uses concepts of topology and geometry and sparks the interest of mathematicians, while physicists are challenged to give a physical interpretation of the solutions to the Dirac equation in the presence of topologically non-trivial, position dependent mass terms. 

Examples of interesting profiles that can replace the homogenous mass term are: kink in one dimension, vortex in two and magnetic monopole in three.

\begin{center}   {\underline{\bf Dirac Equations}}
\end{center}
\vskip 1ex
\centerline{$[\boldsymbol{\alpha}\cdot \mathbf{p} + \beta m] \Psi_E = E \Psi_E$}
\vskip 1ex

\noindent Dirac equation (mass term position-independent, homogenous)\\
Usual:   continuum solutions $E > 0$ and $E<0$\\[1.5ex]
Dirac equation in the presence of a defect \\
 \centerline{(mass term position-dependent)}
\vskip .25ex
\centerline{$[\boldsymbol{\alpha}\cdot \mathbf{p} + \beta\, \phi\, (\mathbf{r})]  \Psi_E = E \psi_E$}

$m \to \phi({\bf r})$: phonon field
\begin{figure}[H]
\begin{center}
\underline{\bf Examples of Defects}\\[.25ex]
\includegraphics[scale=.55]{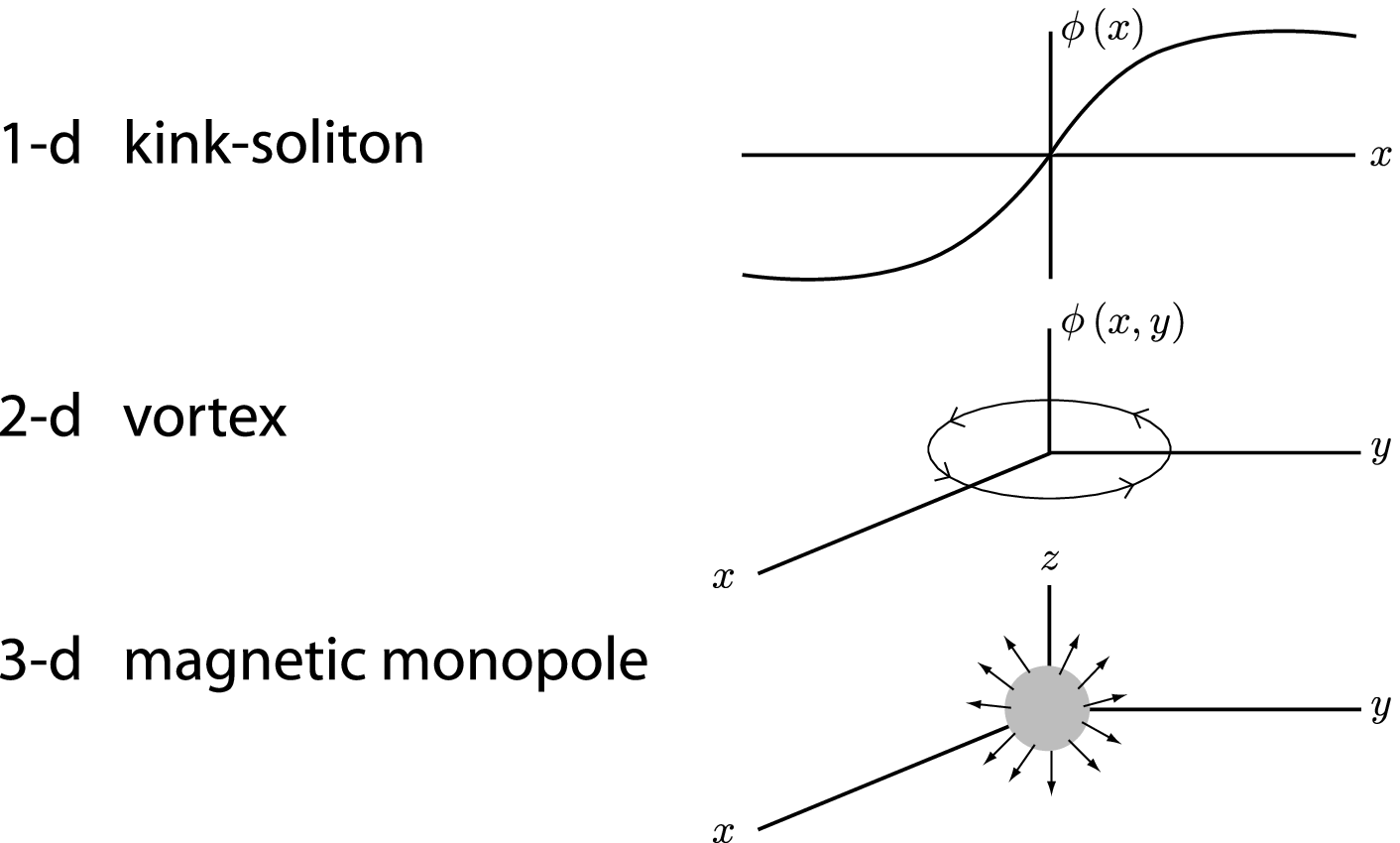}\\[1.75ex]
Continuum solutions $E>0$ and $E<0$ and isolated $E =0$ solution
\end{center}    
\end{figure}

With Rebbi and Rossi we studied Dirac equations in these topologically nontrivial backgrounds. We found that in all instances the spectrum contains the expected positive and negative energy modes, but additionally there are also isolated zero-energy modes. Moreover the zero-energy solutions do not require a specific form for the inhomogenous mass profile; all that matters is that it belongs to a non-trivial topological class. It soon became clear that the presence of the isolated zero-energy modes can be established a priori by mathematical ÒindexÓ theorems, which relate the occurrence of these modes to the topology of the mass term profile and the geometry of the space on which the equation is stated.

This realization delighted us, and encouraged our belief that the mathematical elegance of the zero-mode solution ensures its profound physical significance. The remaining task was to describe the physical role of the zero modes. 
    
\section*{Polyacetylene}

The first and oldest instance of this story concerns polyacetylene. This is a lineal array of carbon atoms, separated by about one angstrom. The dispersion law connecting energy $E$ to momentum $p$ is sketched in the figure. The dashed line indicates the Fermi energy. 
\begin{figure}[H]
\begin{center}
\includegraphics[scale=.7]{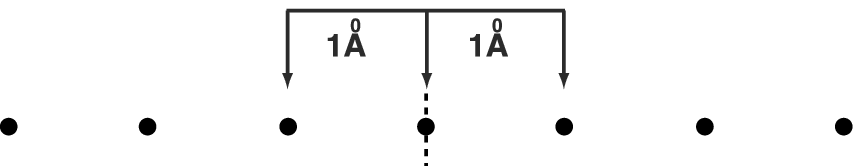}\\[2ex]
\includegraphics[scale=.7]{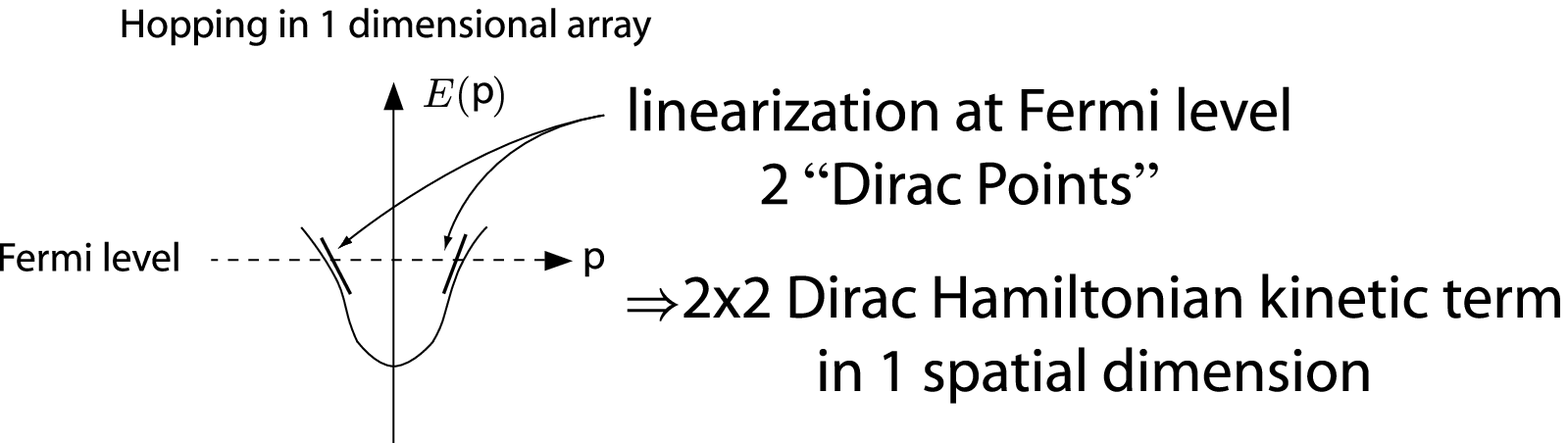}
\end{center}
\end{figure}
The energy curve intersects the Fermi level in two points, called Dirac points. Near these points the dispersion can be linearized. Thus an effective Hamiltonian is a $2 \times 2$ matrix multiplied by $p$, this provides the kinetic term of a one-dimensional Dirac Hamiltonian, but is only half the story. 

It is known that in polyacetylene the uniform array of carbon atoms is unstable.  In fact equilibrium is achieved only after the atoms shift by about .04 angstroms to the left or to the right. Both cases are possible, giving rise to two equivalent domain structures. This is called the Peierls instability, and is modeled in our effective Dirac Hamiltonian by a homogenous mass term that takes either a positive or negative value. 
\begin{figure}[H]
\begin{center}
\includegraphics[scale=.7]{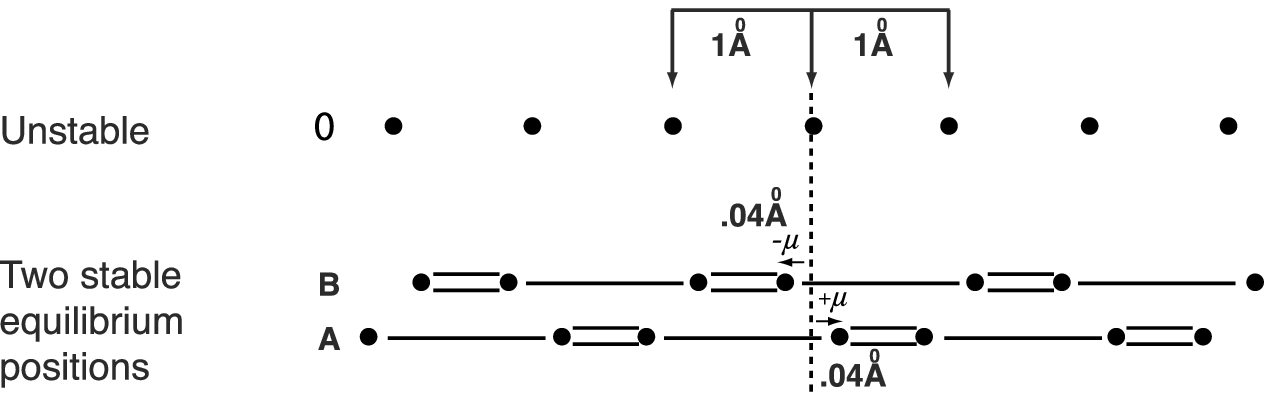}
\end{center}
With these steps we arrive at the Dirac-like equation for polyacetylene.
\end{figure}
\begin{center}
{\underline{\bf Polyacetylene Realization of 1-d Dirac equation}}\\[3ex]

\includegraphics[scale=.6]{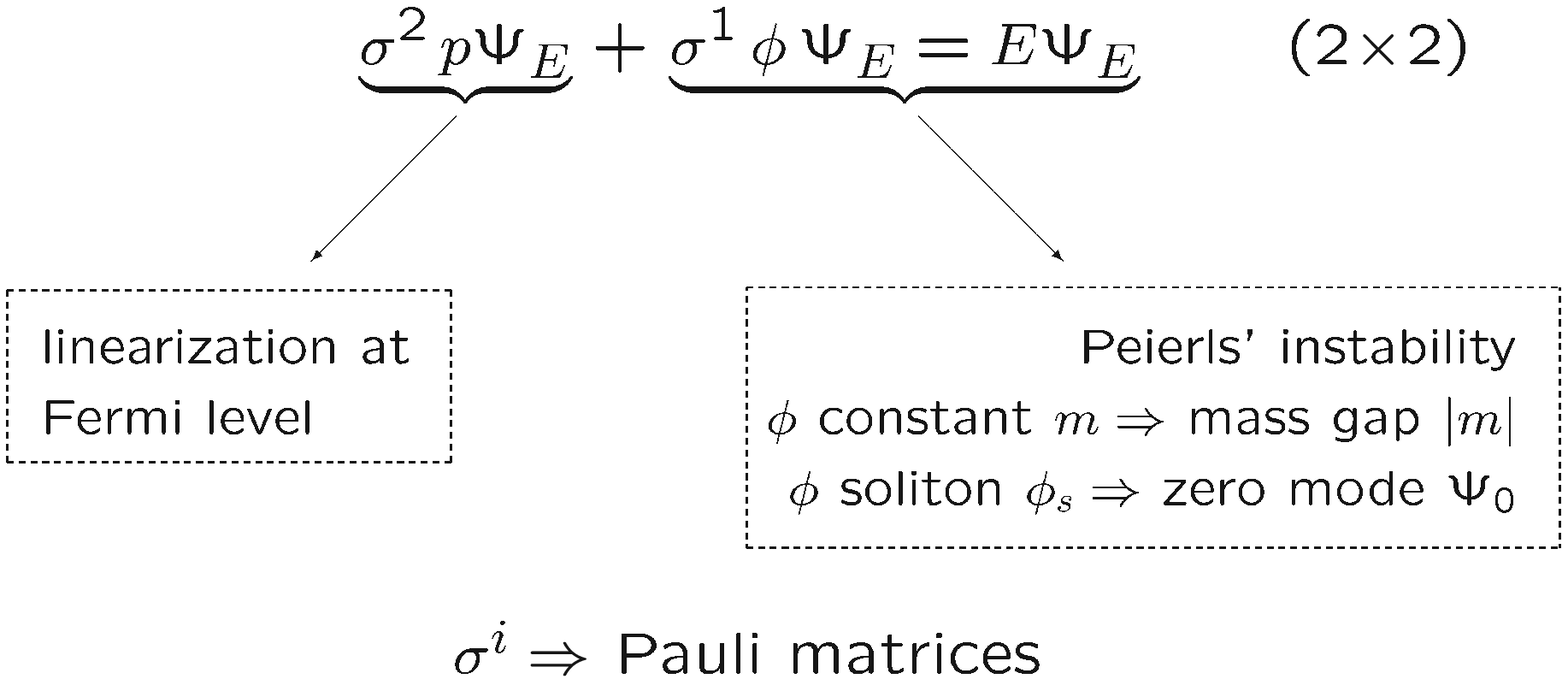}
 \end{center}

When the mass term is inhomogenous  we call it a phonon field, whose non-vanishing value describes the carbon shifts due to the Peierls instability.

The energetics and the profiles of the phonon field are pictured in the figure.
    
  \begin{figure}[H]
   \begin{center}
{\underline{ \bf Energetics of Polyacetylene Phonon Field}}\\[.35em]
 \includegraphics[scale=.7]{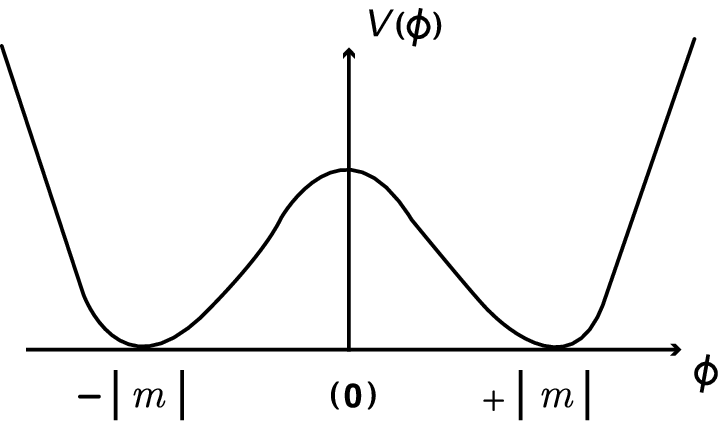}
 \end{center}
{ \small Energy density $V(\phi)$, as a function of a constant phonon field $\phi$. The symmetric stationary point, $\phi = 0$, is  unstable. Stable configurations are at $\phi = + \, | m |, \text{(A) and}\,  \phi = -| m |, \text{(B)}$.}
\end{figure}
 \begin{figure}[H]
\begin{center}
{\underline{\bf Profiles of Polyacetylene Phonon Field}}\\[.35em]
  \includegraphics[scale=.7]{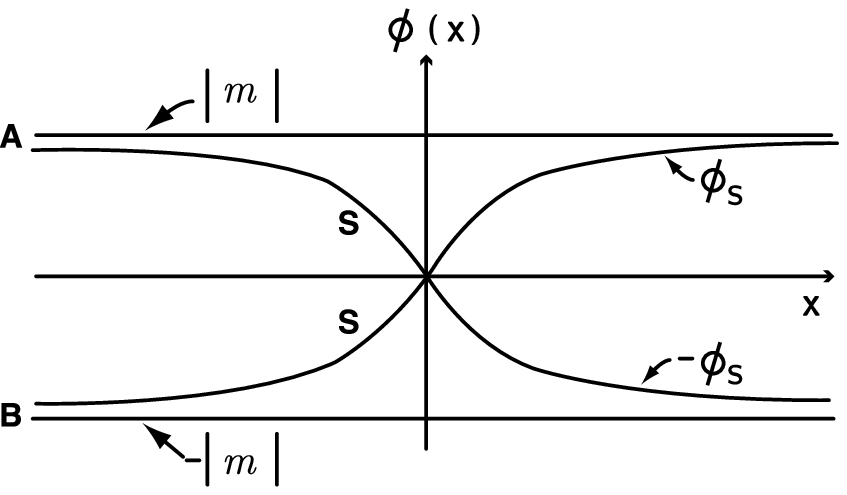}
  \end{center}
  { \small The two constant fields, $\pm \mid m \mid$, correspond to the two domains (A and B). The  kink-soliton fields, $\pm \phi_s$, interpolate between the vacua and represent domain walls.}
\end{figure}

The role of a kink-shaped mass profile is now apparent: it describes a domain wall, interpolating between positive and negative mass terms, that is between the two equivalent domains of polyacetylene. 

A simple calculation shows that no matter what the detailed kink shape might be, the Dirac equation in the presence of a kink always supports a normalizable, zero-energy state. But in fact no explicit calculation is needed because an index theorem guarantees the result.

The question remains how to treat the zero-energy state in the construction of the ground state of the material. Should the state be empty, like the positive energy states, or does it belong to the filled, negative energy states? Dirac has nothing to say about this, because he did not know about zero-energy modes!

\begin{center}
 {\underline{\bf Fractional Charge Q}}
\end{center}

Rebbi and I gave the answer.  Because there is no energy difference between the two possibilities, both have to be taken into account: empty and filled. Moreover the charge Q (or fermion number) of the empty state is $- 1/2$, while the filled state carries charge $+ 1/2$. The fractional number is not an expectation value, but a sharp eigenvalue, that is there are no fluctuations.
 
There are many ways to prove this surprising result. An intuitive argument is based on the fact that our effective Hamiltonian possesses a conjugation symmetry, such that the energy spectrum of filled states is the same as that of empty states, but the charge is opposite. Because of this symmetry our zero energy state lies in the middle of the gap, and the charge difference between empty and filled states is one. Since the charges should also differ only in sign they must take the values $\pm 1/2$.  

In a more formal argument, we construct the charge operator and find\\
\mbox{\hspace{1.5in}   $Q$ = contribution of $E\ne 0$ states $\ +\,  \frac{1}{2}\ (a^\dagger a - a a^\dagger)$}
 
 Here $a^\dagger$   and $a$  are the creation and annihilation operators for the zero mode, obeying the usual fermionic anti-commutation relations, acting on states to fill or empty them.
 
\[
\begin{array}{lll}
a^\dagger a + a a^\dagger &=& 1 \qquad\  aa = 0\\[1ex]
a^\dagger \ket{\text{empty}}  &  = &  \ket{\text{filled}} \\[1ex]
 a  \ket{\text{filled}}&  =  &   \ket{\text{empty}} \\[1ex]
a^\dagger \ket{\text{filled}}  & =  &   0\\[1ex]
a \ket{\text{empty}} & = & 0\\[1ex]
Q \ket{\text{state}} &=& \pm \frac{1}{2} \ket{\text{state}}
\end{array}
\]
It follows that            $Q        =      \pm 1/2$.  

Experimental confirmation came when Schrieffer and collaborators observed that charge fractionalization explains certain novel features of polyacetylene. Thereby they established the physical realty of this unexpected phenomenon. 

As a result physicists opened their minds to the occurrence of charge fractions. Thus when the quantum Hall effect emerged as a fascinating planar quantum system, fractionally charged states helped understand the relevant physics. 

But these fractions arise from a very different mechanism. There is no Dirac equation; there are no topological defects; most significantly, the quantum Hall effect exists in the presence of a magnetic field, which breaks time-inversion invariance , while our Dirac equations preserve time-inversion. 

Therefore, the question arises: beyond one dimension are there physical systems described by an effective Dirac equation that leads to charge fractionalization in a time-inversion symmetric manner. 

Especially challenging is the fact that in higher dimensions the Fermi level defines a surface and its intersection with the energy dispersion is an extended manifold, unlikely to reduce to a finite number of Dirac intersection points. 

But then comes graphene.
\section*{Graphene}

Graphene is a planar array of carbon atoms in a hexagonal lattice. It is a one-atom thick sample of graphite. The hexagon can be displayed as a superposition of two triangular sublattices, each possessing two Dirac points.
     
 \begin{center}
 {\underline{\bf Graphene hexagonal lattice}}
 \end{center} 
 \vspace{-1ex}
 \begin{figure}[h]
 \begin{center}
 \includegraphics[scale=.35]{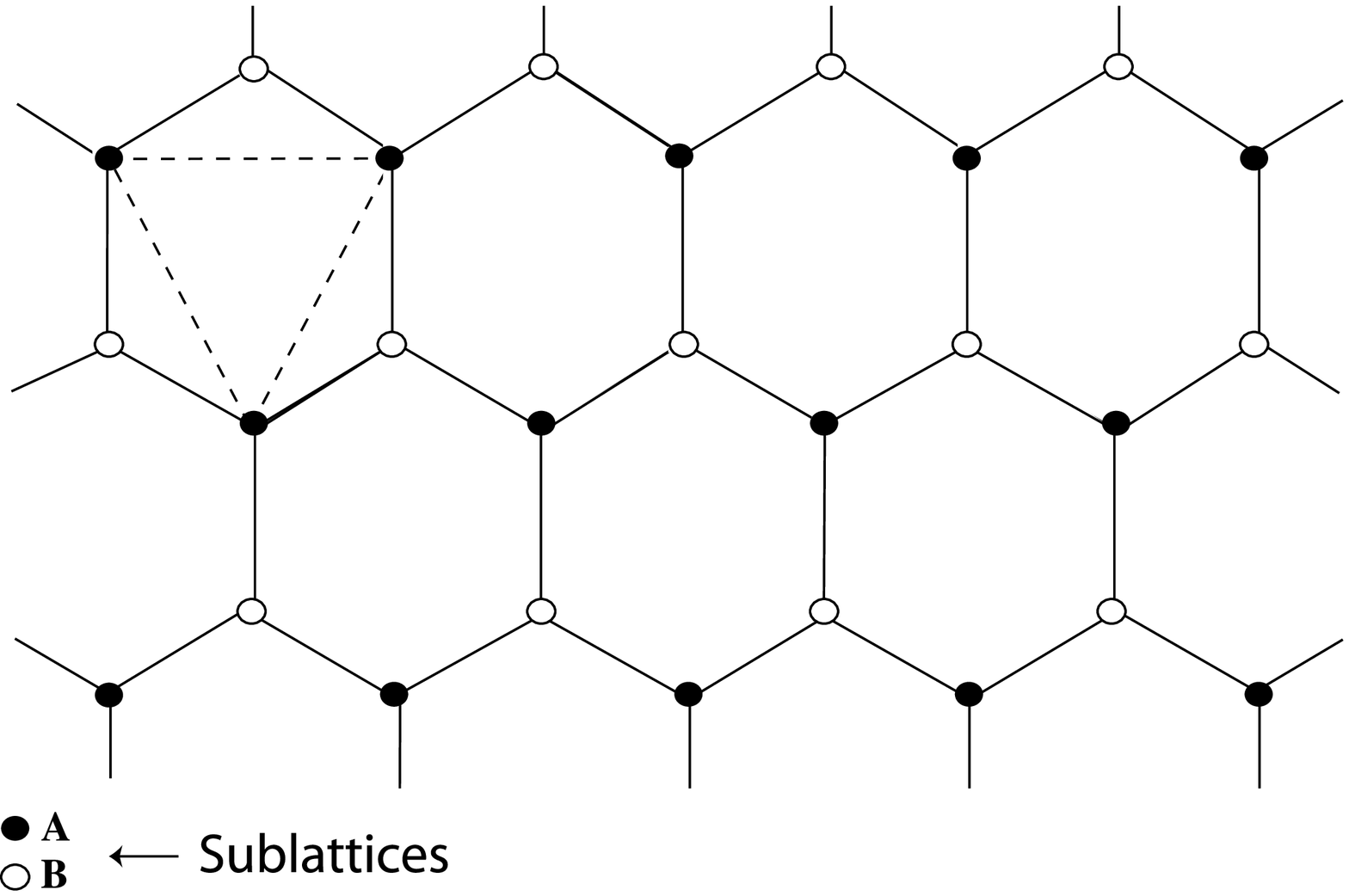}
\end{center}
\end{figure}
Linearization at Fermi level

\noindent 2 ``Dirac points" per sublattice\\[.5ex]


Wallace showed this in 1947. He was a Canadian physicist, Canada's contribution to the American atomic bomb project, where he studied properties of graphite --- an important neutron moderator. After the war work Wallace published his investigation of monolayer graphite, where he demonstrated that due to the specific band structure the energy Ð momentum dispersion can be linearized near two Dirac points, just as in polyacetylene. Almost forty years later, Semenoff --- another Canadian theorist --- showed that the linear approximation leads to a Dirac kinetic term. 

The Dirac structure gives rise to specific experimental effects. These have been observed in graphene, and the 2010 Nobel Prize was awarded to the experimentalists Geim and Novoselov for their graphene work.
     
But this is only half the way to fractional charge. Still needed is a mass-like term, which can take a constant value or an inhomogenous profile in the shape of a vortex.  The constant value would produce a gap in the energy spectrum, while the topologically non-trivial configuration of a vortex would support a zero-energy mode. In other words, what is needed is a planar analog of the Peierls instability. 

Chamon and collaborators have identified for graphene the possibility of such an effect, called a ÒKekul\'{e} structureÓ.  It preserves time-inversion invariance and bonds carbon atoms in the plane with a pattern that generalizes the one-dimensional Peierls instability.     
%
\begin{figure}[H]
 \begin{center}
 \underline{\bf Graphene hexagonal lattice with Kekul\'{e} distortion}
 \includegraphics[scale=.4]{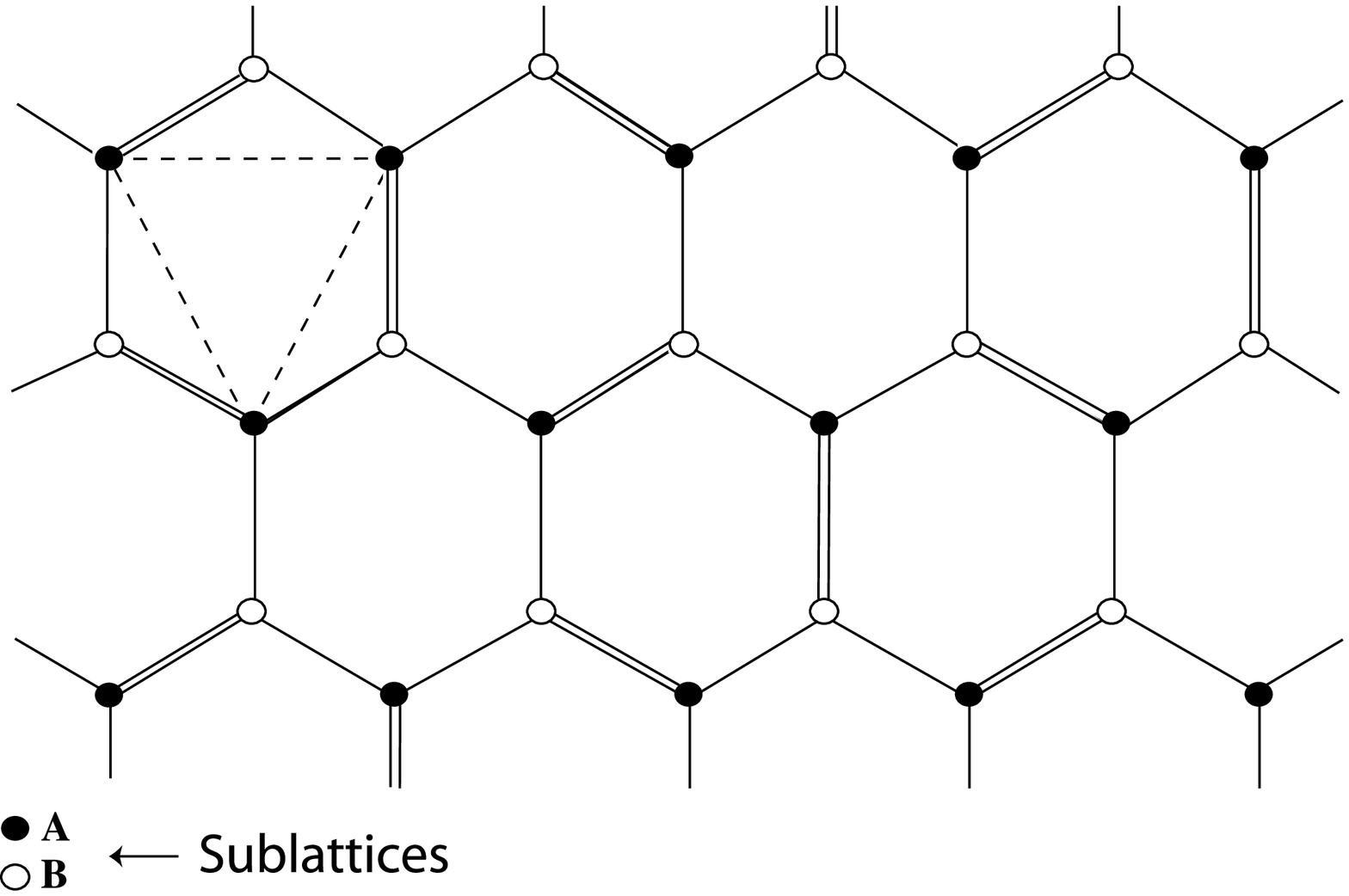}
 \end{center}
\end{figure}
But unlike the Peierls effect there is no guarantee that the Kekul\'{e} structure is always present. If it occurs then the effective Dirac equation reads
\begin{figure}[H]
 \begin{center}
 {\underline{\bf Graphene Realization of (2-d) Dirac Equation}}\\[2ex]
 \includegraphics[scale=.6]{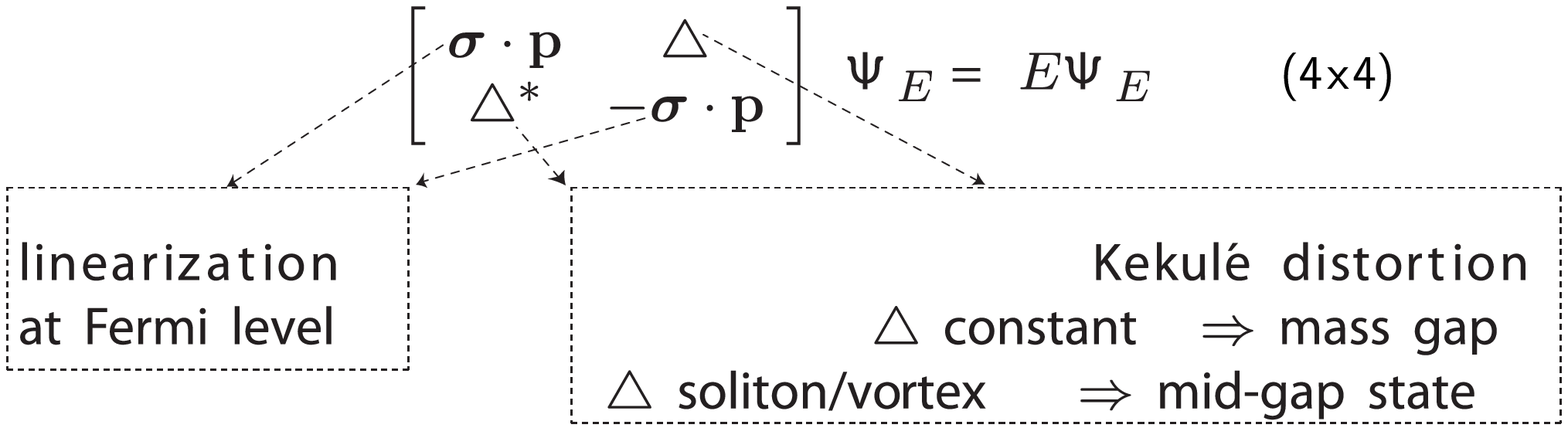}
\end{center}
\end{figure}

With a vortex profile for   $\Delta$ ,  Chamon used the solution to a related equation, which was found earlier by Rossi and me,  to display a zero-energy mode that leads to fractional charge, in the same manner as for one-dimensional polyacetylene.  Thus in spite of the tenuousness of the Kekul\'{e} structure, we have here at the very least an existence proof that fractional charge can also arise in the plane in a time-inversion symmetric scenario.

\section*{Majorana Fermions}

Complex fields describe charged particles and their anti-particles of opposite charge. This is as in the Dirac equation for electrons and positrons. In nature there are also neutral particles that are their own anti-particles. Examples are
      
\begin{itemize}
 \item[]     neutral pion  (spin 0)
   \item[]              photon              (spin 1) 
      \item[]          graviton          (spin 2)
\end{itemize}
All these are integer-spin bosons, and real fields describe them.

The question arises: can there exist neutral half-integer spin fermions that are their own anti-particles?  Thus far experimentalists have not identified any, but theorists like them very much: they use them in speculations on particle physics where they are supersymmetric partners of neutral bosons; in cosmology they possibly constitute dark matter and describe neutrinos; most recently they have arisen in condensed matter physics as exotic superconducting states.
     
Neutral, spin 1/2 particles that are their own anti-particles are called ``Majorana fermions" because the mysterious Italian physicist Majorana discovered the equation that describes them. Let me review. 

\centerline{\underline{\bf Majorana Equation}}
\begin{figure}[H]
\includegraphics{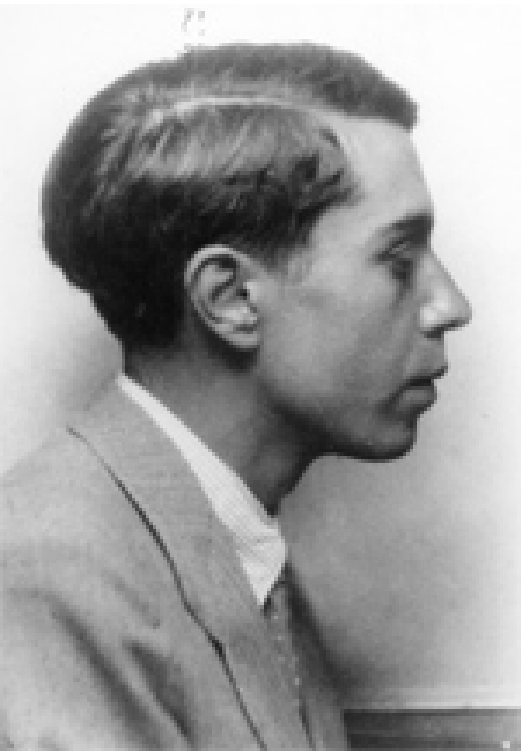}\ \
\begin{minipage}[b][2in][t]{6in}
Majorana Matrix Equation\\[1ex]
$({\boldsymbol \alpha \cdot \bf p} + \beta m) \Psi = i\, \frac{\partial}{\partial t}\, \Psi$\\[1ex]
\indent $\Psi$ real (neutral excitations)\\[1ex]
\mbox{\hspace{.75em} ${\bf p} = \frac{1}{i} \, \nabla$ imaginary}\\[1ex]
\mbox{\hspace{.75em} $\alpha$ real}\\[-1ex]
\mbox{\hspace{.75em} $\beta$ imaginary} \raisebox{1ex}{$\ \big\}$ Majorana Representation} 
\end{minipage} 
\end{figure}
     
Because they are spin 1/2 particles, the Dirac equation remains valid. Majorana realized that there exists a representation of the Dirac matrices  ${\boldsymbol \alpha}$         and  $\beta$       , where   ${\boldsymbol \alpha}$      is real and    $\beta$  is imaginary. In this representation, called the ``Majorana representation," one can demand that the spinor   $\Psi$     is real, and then the Dirac equation becomes the real Majorana equation. This ensures charge neutrality of a spin 1/2 particle, identified with its anti-particle.

The Majorana equation is a further step in the deconstruction of a differential operator:  real second order Laplacian; complex first order Dirac operator; real first order Majorana operator. Thus it possesses mathematical interest in addition to its physical role

In fact matrices arising in the Majorana equation need not possess the above-stated reality properties. Since similarity transformations on matrices do not affect physical content, it is sufficient to demand in any representation that there exists a matrix   $\mathcal{C}$      such that  
\[
\mathcal{C}\, {\boldsymbol \alpha}^\ast\, \mathcal{C}^{-1} = {\boldsymbol \alpha} , \quad \mathcal{C}\, \beta^\ast\, \mathcal{C}^{-1} = -\beta
\]
and the reality condition on $\Psi$ is replaced by the pseudo-reality requirement  $\mathcal{C}\, \Psi^\ast = \Psi$.   In the Majorana representation $\mathcal{C} = I$, while with the usual, complex  ${\boldsymbol \alpha}$      and   $\beta, \
\mathcal{C} {\scriptstyle{=}}\! \left(\begin{array}{cc}0 &  - i\sigma^2    \\ i \sigma^2  & 0 \end{array}\right)$

A four-component spinor satisfying the pseudo-reality condition necessarily takes the form  $ \Psi {\scriptstyle{=}}\!\left(\begin{array}{c}\psi  \\[1ex]i \, \sigma^2\, \psi^\ast \end{array}\right)$ and the four-component equation reduces to the two component Majorana equation.
\[
{\boldsymbol \sigma} \cdot {\boldsymbol {\bf p}}\, \psi + i \, \sigma^2\, \varphi\, \psi^\ast = i\, \frac{\partial}{\partial t}\ \psi \qquad 
\]

Note that the mass term couples    $\psi$      to   $\psi^\ast$. Therefore the equation does not entail charge or particle conservation;  indeed excitations carry no charge.

Recent developments in neutrino particle physics have established that neutrinos have mass. The occurrence of neutrino oscillations indicates that individual species of neutrinos are not conserved. This suggests that neutrinos are Majorana fermions. Nowdays there is intense search for definite verification of the Majorana nature of the neutrino, making the Majorana equation relevant to physics in three dimensions.
      
The Majorana equation taken in two spatial dimensions allows replacing the constant mass term with an inhomogenous vortex profile, whereupon the equation possesses zero modes, as was first found by Rossi and me.

Remarkably this same planar equation has recently been posited by Fu and Kane in their description of a superconductor in contact with a topological insulator. A proximity effect causes Cooper pairs to tunnel through the surface of the topological insulator. This effect is modeled by the planar Majorana equation, which also supports the above mentioned zero mode in the presence of the vortex.

Owing to its Majorana  nature, the zero mode does not carry any charge. Moreover, it entails a novel mode algebra for its mode operators, like the ones introduced earlier in the polyacetylene story.\\
$\centerline{$a a^\dagger + a^\dagger a = 1 \quad $ but $a = a^\dagger$ (Majorana)}\\ \centerline{$ \Rightarrow aa = \frac{1}{2}$}\\$
This novel algebra is neither conventionally fermionic nor bosonic. Its action calls for two states, $\ket{+}, \ket{-}$.
\[
a \ket{+} = \frac{1}{\sqrt{2}} \, \ket{-} \quad a \ket{-} =  \frac{1}{\sqrt{2}}\, \ket{+}
\]

It has been suggested that this algebra will find application in quantum computation. This is because the zero mode is stable against perturbations, since topological considerations guarantee its presence.

Now we wait to learn who will first discover Majorana fermions; particle physicists with neutrino-less double beta decay for example, or cosmologists by identifying dark matter, or condensed matter people with exotic superconductors, or will quantum computers beat out the physicists.

 \section*{Into the Future}

What more is there to do? We have discussed one-dimensional kinks in polyacetylene, two-dimensional vortices in graphene and in exotic superconductors. What about three dimensions, where Dirac originally posited his equation?

With Rebbi, we showed that replacing the constant mass term by a three-dimensional profile of a magnetic monopole again leads to zero-modes in the extended Dirac equation. Since charge remains conserved, this system exhibits fractional charge.

Unfortunately we do not have a good physical candidate for a magnetic monopole, even though material scientists have occasionally discussed effective monopoles, like the ones in spin-ice. But these are not topological defects.
     
Therefore a good future problem would be to determine whether magnetic monopoles exist and whether they induce fractional charge in three dimensions.     

\doeparticle

\end{document}